\documentstyle[12pt,epsf,eqsection,draft]{article}

\begin{document}

\def\spose#1{\hbox to 0pt{#1\hss}}
\def\ltapprox{\mathrel{\spose{\lower 3pt\hbox{$\mathchar"218$}}
 \raise 2.0pt\hbox{$\mathchar"13C$}}}
\def\gtapprox{\mathrel{\spose{\lower 3pt\hbox{$\mathchar"218$}}
 \raise 2.0pt\hbox{$\mathchar"13E$}}}
\def\inapprox{\mathrel{\spose{\lower 3pt\hbox{$\mathchar"218$}}
 \raise 2.0pt\hbox{$\mathchar"232$}}}

 \title{ { \normalsize \hfill \parbox{28mm}{BI-TP 98/33} }\\[40mm]
          \vspace{-2cm} Pseudo-Character Expansions for 
          $U(N)$-Invariant Spin Models on $CP^{N-1}$}

\author{
  \\
  {\small Attilio Cucchieri}                  \\[-0.2cm]
  {\small Tereza Mendes}  \\[-0.2cm]
  {\small\it Fakult\"at f\"ur Physik}       \\[-0.2cm]
  {\small\it Universit\"at Bielefeld}         \\[-0.2cm]
  {\small\it D-33615 Bielefeld, GERMANY}      \\[-0.2cm]
  {\small e-mail: 
       {\tt attilio@physik.uni-bielefeld.de,}} \\[-0.2cm] 
  {\small {\tt mendes@physik.uni-bielefeld.de}} \\[-0.2cm]
  \\[-0.35cm] \and
  {\small Andrea Pelissetto }        \\[-0.2cm]
  {\small\it Dipartimento di Fisica}        \\[-0.2cm]
  {\small\it Universit\`a degli Studi di Pisa}        \\[-0.2cm]
  {\small\it I-56100 Pisa, ITALIA}        \\[-0.2cm]
  {\small e-mail: {\tt pelisset@ibmth.df.unipi.it}}  \\[-0.2cm]
  {\protect\makebox[5in]{\quad}}  
  \\
}

\maketitle
\thispagestyle{empty}   


 \begin{abstract}
We define a set of orthogonal functions on the complex projective 
space $CP^{N-1}$, and compute their Clebsch-Gordan coefficients as
well as a large class of 6--$j$ symbols. We also provide all the 
needed formulae for the generation of high-temperature expansions
for $U(N)$-invariant spin models defined on $CP^{N-1}$.
 \end{abstract}

\vspace{0.5 cm}

\clearpage

\newcommand{\be}{\begin{equation}}
\newcommand{\ee}{\end{equation}}
\newcommand{\<}{\langle}
\renewcommand{\>}{\rangle}
\newcommand{\para}{\|}
\renewcommand{\perp}{\bot}

\def\newhat{\mbox{\v{\ }}}
\def\half{ {{1 \over 2 }}}
\def\smfrac#1#2{{\textstyle\frac{#1}{#2}}}
\def\smhalf{ {\smfrac{1}{2}} }
\def\scra{{\cal A}}
\def\scrc{{\cal C}}
\def\scrd{{\cal D}}
\def\scre{{\cal E}}
\def\scrf{{\cal F}}
\def\scrh{{\cal H}}
\def\scrm{{\cal M}}
\newcommand{\scrmvec}{\vec{\cal M}}
\def\scro{{\cal O}}
\def\scrp{{\cal P}}
\def\scrs{{\cal S}}
\def\scrt{{\cal T}}
\def\ttens{{\stackrel{\leftrightarrow}{T}}}
\def\scrttens{{\stackrel{\leftrightarrow}{\cal T}}}
\def\scrv{{\cal V}}
\def\scrw{{\cal W}}
\def\scry{{\cal Y}}
\def\tauss{\tau_{int,\,\scrm^2}}
\def\taux{\tau_{int,\,{\cal M}^2}}
\newcommand{\taum}{\tau_{int,\,\vec{\cal M}}}
\def\taue{\tau_{int,\,{\cal E}}}
\newcommand{\imag}{\mathop{\rm Im}\nolimits}
\newcommand{\real}{\mathop{\rm Re}\nolimits}
\newcommand{\tr}{\mathop{\rm tr}\nolimits}
\newcommand{\sgn}{\mathop{\rm sgn}\nolimits}
\newcommand{\codim}{\mathop{\rm codim}\nolimits}
\def\textprime{{${}^\prime$}}
\newcommand{\longto}{\longrightarrow}
\def\var{ \hbox{var} }
\newcommand{\gtilde}{ {\widetilde{G}} }
\newcommand{\USp}{ \hbox{\it USp} }
\newcommand{\CP}{ \hbox{\it CP\/} }
\newcommand{\QP}{ \hbox{\it QP\/} }
\def\hboxscript#1{ {\hbox{\scriptsize\em #1}} }

\newcommand{\plotdot}{\makebox(0,0){$\bullet$}}
\newcommand{\plotsmalldot}{\makebox(0,0){{\footnotesize $\bullet$}}}

\def\bsigma{\mbox{\protect\boldmath $\sigma$}}
\def\btau{\mbox{\protect\boldmath $\tau$}}
\def\brho{\mbox{\protect\boldmath $\rho$}}
\def\ba{\mbox{\protect\boldmath $a$}}

\def\bw{\mbox{\protect\boldmath $w$}}
\def\bz{\mbox{\protect\boldmath $z$}}
\def\bv{\mbox{\protect\boldmath $v$}}
\def\bx{\mbox{\protect\boldmath $x$}}
\def\bwbar{\overline{\bw}}
\def\bzbar{\overline{\bz}}
\def\bxbar{\overline{\bx}}
\def\brhobar{\overline{\brho}}
\def\zbar{\overline{z}}
\def\wbar{\overline{w}}

\def\hyper#1#2#3{{Y_{N,#1,#2}^{\hphantom{N,#1}#3}}}
\def\ZZ#1#2#3{{Z_{N,#1,#2}^{\hphantom{N,#1,}#3}}}
\def\Clebsch#1#2#3{{{\cal C}_{N;#1,#2}^{\hphantom{N;#1,}#3}}}

\newcommand{\reff}[1]{(\ref{#1})}

\font\srm=cmr7 		

\newtheorem{theorem}{Theorem}[section]
\newtheorem{corollary}[theorem]{Corollary}
\newtheorem{lemma}[theorem]{Lemma}
\def\proof{\bigskip\par\noindent{\sc Proof.\ }}
\def\qed{\hbox{\hskip 6pt\vrule width6pt height7pt depth1pt \hskip1pt}\bigskip}

\def\kbar{ {\bar{k}} }
\def\lbar{ {\bar{l}} }
\def\mbar{ {\bar{m}} }

%
%
\newenvironment{sarray}{
          \textfont0=\scriptfont0
          \scriptfont0=\scriptscriptfont0
          \textfont1=\scriptfont1
          \scriptfont1=\scriptscriptfont1
          \textfont2=\scriptfont2
          \scriptfont2=\scriptscriptfont2
          \textfont3=\scriptfont3
          \scriptfont3=\scriptscriptfont3
        \renewcommand{\arraystretch}{0.7}
        \begin{array}{l}}{\end{array}}

\newenvironment{scarray}{
          \textfont0=\scriptfont0
          \scriptfont0=\scriptscriptfont0
          \textfont1=\scriptfont1
          \scriptfont1=\scriptscriptfont1
          \textfont2=\scriptfont2
          \scriptfont2=\scriptscriptfont2
          \textfont3=\scriptfont3
          \scriptfont3=\scriptscriptfont3
        \renewcommand{\arraystretch}{0.7}
        \begin{array}{c}}{\end{array}}

\def\zed{{\bf \rm Z}}
\def\szed{{\hbox{\srm Z\kern-.45em\hbox{\srm Z}}}}
\def\R{{\hbox{{\rm I}\kern-.2em\hbox{\rm R}}}}
\def\sR{{\hbox{{\srm I}\kern-.2em\hbox{\srm R}}}}
\def\restrict{ {\mid^{\hskip-0.2em\backslash}} }

\section{Introduction}
\label{sec:In}

Spin systems have been the subject of intense research for a long time.
Indeed they describe the universal features of many different phenomena.
For instance, the $N$-vector model --- a generalization of the Ising 
model ---
describes the critical behavior of dilute polymers, ferromagnets, binary
fluids, liquid helium, and of the Higgs sector of the Standard Model 
at finite temperature \cite{Ma_book,Parisi_book,Zinn-Justin_book}. From 
the point of view of quantum field theory, the $N$-vector model
provides the simplest example for the realization of a non-Abelian global
symmetry. In particular, its two-dimensional version has been extensively 
studied because it shares with four-dimensional gauge theories the property
of being asymptotically free in the weak-coupling perturbative expansion
\cite{Polyakov,BrZJ,Bardeen_76,Kogut}. 

A generalization of the $N$-vector
model is provided by $CP^{N-1}$-models
\cite{DAdda-etal,Witten,Stone,Rabinov,DiVecchia-etal,Rossi}
(for recent work see 
\cite{Campostrini-Rossi-Vicari,Campostrini-Rossi,Plefka-Samuel_97}
and references therein). Their two-dimensional 
version, besides being asymptotically free and invariant under global
$U(N)$-transformations, shows some additional features that are 
expected to hold (though very difficult to prove) in the QCD case.
Indeed, $CP^{N-1}$-models have a local $U(1)$-gauge invariance, 
dynamical appearance of a linear confining potential between 
non-gauge-invariant states, and a non-trivial topological structure with 
instantons, anomalies and $\theta$-vacua. Besides being of interest in
high-energy physics, the $CP^{N-1}$-models have applications 
in condensed-matter physics. Indeed, they describe models that have
a complex vector order parameter (see e.g.\  \cite{Auerbach}).

Spin systems can be studied in a variety of ways: predictions
can be obtained using field-theory methods --- based 
essentially on perturbation theory \cite{Zinn-Justin_book} ---
or using numerical techniques, e.g.\ 
Monte Carlo simulations or extrapolations
of high-temperature series. The latter method (see for 
instance \cite{Guttmann})
has proved very powerful  in providing extremely accurate estimates of 
critical parameters. For spin systems, such as the 
$N$-vector model, precise results \cite{noi-HT-Ngt3,Butera-Comi-2d}
are obtained even in the case of asymptotically free models, for which 
$\beta_c = +\infty\,$.

Various methods have been used to generate high-temperature expansions 
for the $N$-vector model (for a review, see \cite{Stanley_DG,McKenzie}).
The most straightforward method is the cumulant approach, or the closely
related finite-cluster approach 
\cite{Stephenson-Wood_68,Stanley_67,Joyce-Bowers_66}. 
An improved method is the star-graph expansion, in which only star 
graphs need to be generated
\cite{Domb_DG}. A further improvement
was originally introduced by Joyce \cite{Joyce}. He proposes to expand 
the interaction term in hyperspherical harmonics. This allows a 
fast computation of the partition function for all star graphs. The main 
difficulty of the method is the computation of the Clebsch-Gordan
coefficients, 6--$j$ symbols and higher-order group invariants. 
Domb and coworkers \cite{Domb72,Domb76,Domb79} developed an
algebraic method to compute the group factors associated to each graph.
In Ref.\  \cite{FSS_1d} explicit expressions for the Clebsch-Gordan 
and the 6--$j$ symbols were presented. Making use of simple algebraic rules
\cite{noi-Melbourne} these general results allow the determination of 
all the group factors that are needed in the generation of high-temperature
series of present-day length. These techniques have been recently used to 
generate long series in two and three dimensions 
\cite{noi-HT-Ngt3,noi-Melbourne,noi-HT}. We should mention that 
there exists another method that is extremely efficient in the 
generation of high-temperature series, the linked-cluster technique
\cite{Wortis-etal_69,Wortis,Luscher-Weisz,Butera-Comi}, which is 
an expansion in terms of free graphs.

In this paper we will generalize the results of Ref.\  \cite{FSS_1d} to
$CP^{N-1}$-models. We will use a representation
of the hyperspherical harmonics\footnote{These functions generalize
the usual spherical harmonics for the three-dimensional
rotation group \cite{Edmonds}. Such a generalization is well-known
for any symmetric space \cite{Helgason_2,Helgason,Takeuchi}.} in terms
of completely symmetric and traceless tensors. The computation of the 
Clebsch-Gordan coefficients is then
a straightforward combinatoric exercise. 
Moreover, using the same technique, we are able to compute a very large
class of 6--$j$ symbols, essentially all symbols that are needed
in the generation of high-temperature series of present-day length.

The paper is organized as follows: in Sect.\ 2 we introduce the hyperspherical
harmonics on $CP^{N-1}$ and discuss some general properties.
General results and formulae for Clebsch-Gordan coefficients and 
6--$j$ symbols are reported in Sect.\ 3. In Sect.\ 4 we discuss the 
applications and report the high-temperature expansions of the 
Gibbs weight for the 
Hamiltonians that are used in studies of $CP^{N-1}$ models.
The derivations of the results  are reported in the Appendix.

\section{$CP^{N-1}$-Hyperspherical Harmonics}
\label{sec:HH} \label{sec2}
In this section we introduce a class of orthogonal
functions, which we call hyperspherical harmonics, defined on $CP^{N-1}$.
These functions will provide the basis for expanding the Gibbs weight
for $U(N)$-invariant spin models.

Let us first define $CP^{N-1}$. Its usual definition is the following:
consider $\bw,\bz\in C^{N} - \{0\}$ and introduce the following equivalence
relation:
\begin{itemize}
\item[[Rel1]] $\bw\sim \bz$ if and only if there exists $a\in C$ such that 
$\bw = a \bz$.
\end{itemize}
Then
\be
CP^{N-1} = (C^N - \{0\})/ \sim\; .
\ee
A second definition is the following: consider
vectors $\bw$ belonging to the $(2N-1)$-dimensional complex sphere,
i.e.\  $\bw\in C^N$ such that 
$\overline{\bw}\cdot \bw = 1$, and introduce the equivalence relation:
\begin{itemize}
\item[[Rel2]] $\bw\sim \bz$ if and only if there exists $a\in C$, $|a|=1$, 
 such that $\bw = a\bz$.
\end{itemize}
Then
\be
CP^{N-1} = \{ \bz \in C^N: \overline{\bz} \cdot\bz = 1\} / \sim\; .
\label{CPdef2}
\ee
This second definition is obviously equivalent to the previous one. In
our work we will use this representation.

We mention a third definition that immediately shows that 
$CP^{N-1}$ is a symmetric space of rank one 
(see e.g.\  Refs.\  \cite{Helgason_2,Cartan}). 
Given $\brho=(1,0,\ldots,0)$, notice that any $\bz\in C^N$ such 
that $\overline{\bz}\cdot\bz = 1$ can be written as 
$\bz = U\brho$, with $U\in U(N)$. This representation is not unique:
if $U,V\in U(N)$ are such that $\bz = U\brho = V\brho$, then we can
write $\brho = U^+ V \brho\,$. It is trivial to see that the set of 
$R\in U(N)$ such that $\brho = R\brho$ is isomorphic to $U(N-1)$.
Therefore we obtain the result
\be
\{\bz \in C^N: \overline{\bz} \cdot \bz = 1\} \simeq 
U(N)/U(N-1)\;.
\ee
It follows easily
\be
CP^{N-1} = SU(N)/S(U(N-1)\times U(1))\,.
\ee
It is interesting to observe that for $N=2$, $CP^1\simeq S^2$.
The mapping is obtained as follows. Given $\bz\in C^2$ such that 
$\overline{\bz} \cdot \bz = 1$, consider\footnote{
In this paper we will use the usual summation convention over 
repeated indices.}
\be
s^i \,\equiv\, \overline{z}_\alpha 
(\sigma^i)^{\alpha}_{\hphantom{\alpha}\beta} z^\beta,
\label{CP1S2}
\ee
where $\sigma^i$ are the Pauli matrices. Using the completeness
relation
\be
(\sigma^i)^{\alpha}_{\hphantom{\alpha}\beta} \,
(\sigma^i)^{\gamma}_{\hphantom{\gamma}\delta} \; =  \;
2\, \delta^\alpha_\delta \delta^\gamma_\beta\, - \,
  \delta^\alpha_\beta \delta^\gamma_\delta,
\label{Paulicomplete}
\ee
it is easy to verify that $s^i s^i = 1$, i.e.\  $s\in S^2$. 
Let us now show that the mapping is bijective. Given 
$s\in S^2$, if $t = (0,0,1)$, there exists $R\in SO(3)$, 
such that $R^{ij} t^j = R^{i3} = s^i$. Now given $R\in SO(3)$, it is well
known that there exists $U\in SU(2)$ such that 
$(U^+ \sigma^i U)^{\alpha}_{\hphantom{\alpha}\beta} = R^{ij}
(\sigma^j)^{\alpha}_{\hphantom{\alpha}\beta}\,$. Let us consider
$z^\alpha \equiv U^{\alpha}_{\hphantom{\alpha}\beta} x^\beta$, with
$\bx = (1,0)$. Substituting this expression for $\bz$ in Eq.\ 
\reff{CP1S2}, it is easy to see that one reobtains $s^i = R^{i3}$.
Finally, suppose that $\bz_1$ and $\bz_2$ are 
unimodular and both satisfy Eq.\  \reff{CP1S2} with the same 
$s^i$. Using Eq.\ \reff{Paulicomplete} it is then easy to verify that 
\be
|\overline{\bz}_1\cdot \bz_2|^2 =\, {1\over2} (1 + s^i s^i)\, =\, 1 = \, 
   |\bz_1|^2 |\bz_2|^2.
\ee
This implies $\bz_1 = e^{i\alpha} \bz_2$, for some $\alpha \in \Re\,$.
Thus, $\bz_1$ and $\bz_2$ are equivalent in $CP^{N-1}$.

Let us now define functions on $CP^{N-1}$. It is obvious 
that a function $f$ on $CP^{N-1}$ 
(defined in Eq.\  \reff{CPdef2}) extends naturally to a function 
$\widehat{f}$ on the complex sphere. Indeed, it is enough to define
$\widehat{f}(\bz) = f(\{\bz\})$, where $\{\bz\}$ is the equivalence
class (defined by [Rel2]) to which $\bz$ belongs. On the other 
hand, a given function $f$ on the complex sphere defines a function
on $CP^{N-1}$ if and only if $f(\bz) = f(\bw)$ for $\bz\sim\bw$, 
or --- in physicists' language --- if $f(\bz)$ is invariant under the 
$U(1)$-gauge transformations $\bz\to e^{i\alpha} \bz$, i.e.\ 
\be
f(e^{i\alpha} \bz, e^{-i\alpha} \overline{\bz}) = 
f(\bz,\overline{\bz}),
\ee
for any $\alpha$. In the following we will identify the space of 
functions on $CP^{N-1}$ with the space of $U(1)$-gauge-invariant
functions defined on the complex $(2 N - 1)$-dimensional sphere.
We also introduce a normalized $U(N)$-invariant measure 
\be
d\Omega(\bz,\overline{\bz}) = 
{2^{-N}\over {\cal S}_{2N}}\
d\overline{\bz}d\bz\ \delta(\overline{\bz}\cdot\bz - 1)\; ,
\ee
where ${\cal S}_{2N}$ is the surface area of the sphere $S^{2N-1}$
\be
{\cal S}_{2N} = {2 \pi^N\over (N-1)!}\;.
\ee
Finally, we consider the unitary representation $T(R)$ of $SU(N)$ on 
$L^2(CP^{N-1})$ defined by 
$(T(R)f)(\bz) = f(R^{-1}\bz)$. 
We want to find an orthogonal Hilbert-space decomposition of
$L^2(CP^{N-1})$ into subspaces such that the
representation $T(R)$ restricted to each subspace is
irreducible.  The needed decomposition\footnote{See e.g.\  
Refs.\  \cite{Helgason_2,Helgason,Takeuchi,Barut}.}  
turns out to be precisely the
decomposition of $L^2 ( CP^{N-1} )$ into eigenspaces
of the {\em Laplace-Beltrami operator}
${\cal L} = {\cal L}_{CP^{N-1}}$.\footnote{
   The Laplace-Beltrami operator on $CP^{N-1}$ can be defined as 
   the Casimir operator of the unitary representation of $SU(N)$ 
   on $L^2(CP^{N-1})$. See e.g.\  Ref.\  \cite{Barut}.
}
In fact, it can be
proved\footnote{See~\cite{Takeuchi}, Theorem 13.1, p.\ 231 and 
Corollary 14.4, p.\ 243.}
that:
\begin{itemize}
\item[(a)] The eigenvalues\footnote{Note that our ${\cal L}$ is
the {\em negative}\/ of the usual Laplacian, i.e.\ it is a
{\em positive}\/-semidefinite operator.} 
            of ${\cal L}$ are 
\be
\lambda_{N,k}   \;=\;   k \, (N + k - 1)
                \,\ge\, 0 \; \mbox{,}
\label{lambda}
\ee
where  $k = 0, 1, 2, \ldots \;$.
The corresponding eigenspace $E_{N,k}$ has 
dimension\footnote{
   For a proof see 
   Appendix \ref{app:HH} below. Note that
   ${\cal N}_{N,0}= 1$ and ${\cal N}_{N,1}=N^2-1$ for all $N$.
}
\be
   {\cal N}_{N,k}  \;\equiv\;  \dim \, E_{N,k} \; \, = \; \, 
   (N-1) (N-1+2k) \left[ {(N+k-2)! \over k! (N-1)!}\right]^2,
\label{eq:dim}
\ee
and can be given several equivalent descriptions:
     \begin{itemize}
     \item[(i)] $E_{N,k}$ consists of the restrictions to the 
                complex sphere $\overline{\bz}\cdot\bz=1$ of the 
                $U(1)$-gauge invariant
                polynomials $P_k(\bz,\overline{\bz})$
                that are homogeneous of degree $k$ 
                in $\bz,\overline{\bz}\in C^N$  and
                that satisfy Laplace's equation
                $\partial\overline{\partial} P_k = 0$ in $C^N$.
     \item[(ii)] Consider the tensors
                  $\hyper{k}{\beta_1\ldots\beta_k}{\alpha_1\ldots\alpha_k}
                  (\bz)$
		  of rank $k$ in $\bz$ and $\bzbar$ that are 
                  {\em completely symmetric} under exchange of the
                  $\alpha$ and $\beta$ indices and that are 
                  {\em traceless}, i.e.\ vanish 
                  under any contraction of an $\alpha$-index with a
                  $\beta$-index: $\hyper{k}{\gamma\beta_2\ldots\beta_k}%
{\gamma\alpha_2\ldots\alpha_k} (\bz) = 0\,$.
                  We call these tensors spin-$k$ hyperspherical
                  harmonics (they are described in more detail below).
                  $E_{N,k}$ is then spanned by the tensors
                  $\hyper{k}{\beta_1\ldots\beta_k}{\alpha_1\ldots\alpha_k}$
                  as the indices range over the $N^{2k}$ allowable
		  values.
     \end{itemize}
\item[(b)] Each eigenspace $E_{N,k}$ is left invariant by
	   $T ( R )$. Moreover, the
           representation $T ( R ) \restrict E_{N,k}$
           of $SU(N)$ is irreducible.
\item[(c)] $L^2( CP^{N-1} ) \; = \;
	   \bigoplus\limits_{k=0}^{\infty} \; E_{N,k} \;$ (orthogonal Hilbert
           space decomposition).
\end{itemize}
	
\vskip 0.3cm

To make all this concrete, we can write:
\be
\hyper{k}{\beta_1\ldots\beta_k}{\alpha_{1} \ldots \alpha_{k}}
        (\bz) \; \, \equiv \; \,
\mu_{N,k} \; \left[  z^{ \alpha_{1} } \ldots z^{ \alpha_{k} }
              \overline{z}_{\beta_1} \ldots \overline{z}_{\beta_k} 
       - \mbox{Traces}  \right]\; ,
\label{eq:Ydef}
\ee
where $\bz \in C^{N}$ with $\bzbar\cdot\bz=1$,  ``Traces'' is such that
    $Y_{N,k}$
   is completely symmetric and traceless and
\be
\mu_{N,k} \; = \;
  {1\over k!} \left[ {(N-1+2k)!\over (N-1)!}\right]^{1/2}.
\label{eq:mu}
\ee
Explicit examples are
[the general formula is given in equation \reff{Ygeneral}] :
\begin{eqnarray}
Y_{N,0}(\bz) & = & 1; \\[4mm]
\hyper{1}{\beta}{\alpha}(\bz) & = & \sqrt{N(N+1)} 
\left(z^{\alpha} \overline{z}_\beta - {1\over N} \delta^\alpha_\beta\right);
     \\[4mm]
\hyper{2}{\gamma \delta}{\alpha\beta}(\bz) & = &
    {1\over2}\,	\sqrt{ N(N+1)(N+2)(N+3)}\nonumber \\
   & &  \quad \times 
   \left[ z^{\alpha} z^{\beta} \overline{z}_{\gamma} \overline{z}_{\delta} 
	 -  \frac{1}{N+2} \left( \delta^{\alpha}_{\gamma}
			z^{\beta} \overline{z}_{\delta}
	+ \; 3 \; \mbox{permutations}  \right)  \right.  \nonumber\\
 & & \qquad \quad  + \left. \frac{1}{(N+1)(N+2)} \left( 
                \delta^{\alpha}_{\gamma} \delta^{\beta}_{ \delta} +
                \delta^{\alpha}_{ \delta} \delta^{\beta}_{ \gamma} 
               \right) \right]\; .
\end{eqnarray}
We note that for $N = 2$ we have $\lambda_{2,k} = k(k+1)$ and 
${\cal N}_{2,k} = 2 k + 1$, as expected since $CP^1 \simeq S^2$.
Moreover, using the mapping \reff{CP1S2},
one can show that the $Y$'s are linear combinations of the usual
spherical harmonics. They can also be related to the real-valued
hyperspherical harmonics $(Y_{{\rm sphere}})_{N,k}^{i_1\ldots i_k}$ 
introduced in Ref.\ \cite{FSS_1d} by 
\be
(Y_{{\rm sphere}})_{3,k}^{i_1\ldots i_k}(s) \;=\; 
 2^{-k/2} Y^{\hphantom{2,k}(\alpha)_k}_{2,k,(\beta)_k} (\bz)
(\sigma^{i_1})^{\beta_1}_{\hphantom{\beta_1}{\alpha_1}} \ldots
(\sigma^{i_k})^{\beta_k}_{\hphantom{\beta_k}{\alpha_k}} \; ,
\ee
where $s$ is related to $\bz$ by Eq.\  \reff{CP1S2} and we have 
used multiindex notation $(\alpha)_k = \alpha_1\ldots\alpha_k\,$.
For $N = 2$ we can also prove the identity
\be
Y_{2,k,(\beta)_k}^{\hphantom{2,k}(\alpha)_k} (\bz)\; =\; 
(-1)^k \; Y_{2,k,(\beta)_k}^{\hphantom{2,k}(\alpha)_k} (\bw),
\label{simmetriaN=2}
\ee
where $\bw$ is defined by: $w^\alpha = \epsilon^{\alpha\beta} \zbar_\beta$,
$\overline{w}_\alpha = \epsilon_{\alpha\beta} z^\beta$. Here
$\epsilon_{\alpha\beta} = \epsilon^{\alpha\beta}$ is the completely
antisymmetric tensor satisfying $\epsilon^{12} = \epsilon_{12} = 1$.
In group-theoretical terms, Formula \reff{simmetriaN=2} is related
to the fact that $SU(2)$-representations are self-conjugate.

The normalization $\mu_{N,k}$ is chosen so that the following 
orthogonality relation holds (see Appendix \ref{app:HH}):
\be
\int d\Omega(\bz,\bzbar) \; \;
        \hyper{k}{(\beta)_k}{(\alpha)_{k}}(\bz) \;
        \hyper{l}{(\delta)_l}{(\gamma)_l}(\bz) \; \, = \; \,
      \delta_{k l} \; \,
	I_{N,k,(\beta)_k;(\delta)_k}^{\hphantom{N,k}(\alpha)_k;(\gamma)_k}
   \;,
\label{eq:ortho}
\ee
where $I_{N,k}$ is the ``identity" in the space of spin-$k$ 
symmetric and traceless tensors, i.e.\   the unique orthogonal 
projector onto the space of completely symmetric and traceless tensors
of rank $k$.
The trace of this operator is given by 
\be
I_{N,k,(\beta)_k;(\alpha)_k}^{\hphantom{N,k}(\alpha)_k;(\beta)_k} = \; 
 {\cal N}_{N,k}   \;\equiv\; \dim \, E_{N,k}   \;,
\label{eq:trace}
\ee
as of course it must be.
We remark that 
$Y_{N,k}(\bz) \cdot Y_{N,k}(\bz) \,\equiv\,
   \hyper{k}{(\beta)_k}{(\alpha)_k} (\bz)
   \hyper{k}{(\alpha)_k}{(\beta)_k} (\bz)$
is independent of $\bz$
[by $U(N)$ invariance], and hence 
\be
Y_{N,k}(\bz) \cdot Y_{N,k}(\bz) \; = \;
{\cal N}_{N,k}
\label{YdotY}
\ee
by \reff{eq:ortho} and \reff{eq:trace}.

As stated in the theorem given at the beginning of this section, the 
hyperspherical harmonics are a complete set of functions 
on $L^{2}(CP^{N-1})$. 
Thus any $U(1)$-{\em gauge-invariant} 
function $f(\bz)$ can be expanded as 
\be
f(\bz) \; =\; \sum_{k = 0}^{\infty}
    \;  \widetilde{f}_{k,(\beta)_k}^{\phantom{k,}(\alpha)_k} \;
     \hyper{k}{(\alpha)_k}{(\beta)_k} (\bz)\;,
\label{genericfexp}
\ee
where
\be
 \widetilde{f}_{k,(\beta)_k}^{\phantom{k,}(\alpha)_k} \; = \;
\int d\Omega(\bz,\bzbar)\; \, f(\bz) \; 
     \hyper{k}{(\beta)_k}{(\alpha)_k} (\bz)
\; \mbox{.}
\label{tildef}
\ee
For smooth functions this expansion converges very fast. Indeed, 
if $f(\bz)$ is infinitely differentiable, then as $k\to\infty$ the 
coefficients of the expansion go to zero faster than any inverse power of $k$
(see Appendix \ref{app:Exp}).

The completeness of the hyperspherical harmonics can be expressed through
the relation\footnote{Note that the normalization here follows
directly from the one defined for (\ref{eq:ortho}).}
\be
\sum_{k = 0}^{\infty} \; 
     \hyper{k}{(\alpha)_k}{(\beta)_k} (\bz)\,
     \hyper{k}{(\beta)_k}{(\alpha)_k} (\bw)\;
     =\; \delta(\bz,\bw)
\label{complrel}
\ee
where the $\delta$-function is defined with respect to the measure 
$d\Omega(\bz,\bzbar)$.

Finally, let us consider a $U(N)$-invariant function $f$ defined on 
$CP^{N-1}$. Let $f$ be a function of {\em two}\/ ``spins''
$\bz \mbox{,} \; \bw$, i.e.\ a function 
of $|\bzbar\cdot\bw|$.  We want now to compute its expansion in 
terms of hyperspherical harmonics.
Using Schur's lemma we can write
\be
f(|\bzbar \cdot \bw| ) \; = \;
	\sum_{k=0}^{\infty} \; F_{N,k} \; \, Y_{N,k}(\bz) \cdot
                       Y_{N,k}(\bw)
\; \mbox{.}
\label{eq:exp}
\ee
The ``Traces'' terms can be dropped from either one of the $Y$'s
in the scalar product above, since the other $Y$ is traceless. Also,
since the scalar product is $U(N)$-invariant,
we can rotate $\bz$ to ${\brho} = \left( 1\mbox{,}\; 0\mbox{,}\;
\ldots\mbox{,} \; 0\right)$, and correspondingly rotate $\bw$ to some
$\bv$ with $|\bzbar \cdot \bw | = |\brhobar \cdot \bv| = |v^1|$. In this
way we obtain
\be
Y_{N,k}(\bz) \cdot Y_{N,k}(\bw)  =  \;
Y_{N,k}({\brho}) \cdot Y_{N,k}(\bv) \;
 =  \; \mu_{N,k} \;
\hyper{k}{1\ldots 1}{1\ldots 1}(\bv)
\; \mbox{.}
\label{Y_dot_Y}
\ee
Now $\hyper{k}{1\ldots 1}{1\ldots 1} (\bv)$ can be expressed in terms of
{\em Jacobi polynomials}\/\footnote{This expression corresponds to the 
relation between $Y_{l0}$ and Legendre polynomials for the usual
spherical harmonics. For definitions and properties for Jacobi
polynomials see~\cite{GR}, pp.\ 1035--1037.}
as (see Appendix \ref{app:HH})
\be
\hyper{k}{1\ldots 1}{1\ldots 1}(\bv) \; =\; 
{{\cal N}_{N,k}\over \mu_{N,k}}\,
{P_k^{(N-2,0)} (2 |v^1|^2 - 1)\over P_k^{(N-2,0)} (1)}\; ,
\label{YGegen}
\ee
and therefore
\be
Y_{N,k}(\bz) \cdot Y_{N,k}(\bw) \; = \;
{\cal N}_{N,k} \,
{P_k^{(N-2,0)} (2|\bzbar \cdot \bw|^2 - 1) \over P_k^{(N-2,0)} (1)}
\;.
\label{YYGeg}
\ee
If we now invert Eq.\  \reff{eq:exp} using the orthogonality relation
\reff{eq:ortho} and Eq.\ \reff{YdotY} we get
\be
F_{N,k} \; = \;  \int d\Omega(\bz,\bzbar) \; \, 
f(|\bzbar \cdot \bw| ) \; {Y_{N,k}(\bz) \cdot Y_{N,k}(\bw) \over
                          {\cal N}_{N,k}} \; \mbox{.}
\ee
This can be rewritten, using the $U(N)$-invariance of the
measure and Eq.\ \reff{YYGeg}, as
\be
F_{N,k} \; = \;  \int d\Omega(\bz,\bzbar) \; \,
                        f(|z^1| ) \;
                {P^{(N-2,0)}_k(2|z^1|^2-1)  \over
                 P^{(N-2,0)}_k(1) }
\label{eq:coef2}
\; \mbox{.}
\ee
Now the integrand depends only on $z^1$ and we can integrate out
the other coordinates. 
We finally  obtain
\be
F_{N,k} \; = \; (N-1) \int^1_{0} dt\;\; (1-t)^{N-2}\;
        f(\sqrt{t})\; {P_{k}^{(N-2,0)}(2t-1)\over  P_{k}^{(N-2,0)}(1)}
\label{Fgeneralformula}
\; \mbox{.}
\ee
{}From the general properties of the hyperspherical harmonics we can derive 
the following properties of the coefficients $F_{N,k}$ (for the proofs of
properties 1 and 2,
see Appendix \ref{app:Exp}):
\begin{enumerate}
\item If $f(t)$ is positive\footnote{More precisely, it suffices that
      $f$ be {\em nonnegative}\/ and {\em not almost-everywhere-vanishing}\/.}
      for $t\in [0,1]$, then 
      $|F_{N,k}| < F_{N,0}$ for all $k \neq 0$. 
\item If $f(\sqrt{t})$ is smooth (i.e.\ $C^{\infty}$),
      then $\; \lim\limits_{k \to \infty} k^n F_{N,k} = 0$ for every $n$.
\item If $f(t) = t^{2l}$, then the integral in~\reff{Fgeneralformula} 
      can be performed explicitly
      and the coefficients $F_{N,k}$ are given by\footnote{
      See Eq.\  11, p.\ 583, vol.\ 2 of Ref.\  \cite{Prudnikov}.
      This result can also be obtained using Formula 7.391.3 of \cite{GR},
      once one corrects the following misprint: 
      $\Gamma(\alpha+1)$ in the numerator should be replaced by
      $\Gamma(n+\alpha+1)$.}
       
\be
F_{N,k}^{\left(l\right)} \; = \; \cases{ 
\displaystyle {(N-1)!\, \left( l! \right)^2 \over 
              (N-1+l+k)! \, (l-k)!} \quad &
   if $k \leq l$ \cr
     \noalign{\vskip 8pt}
   0 \quad & otherwise \cr }
\ee
and are, in particular, always nonnegative.
It immediately follows that for a generic function of the form
\be
f(t) \; = \; \sum_{l = 0}^{\infty} \;
     f_{l} \, t^{2l}
\; \mbox{,}
\ee
the coefficients $F_{N,k}$ are given by
\be
F_{N,k} \; = \; \sum_{l = k}^{\infty} \; f_{l} \;
          F_{N,k}^{\left(l\right)}
\label{FTaylor}
\; \mbox{.}
\ee
Therefore, if all the coefficients $f_{l}$ are nonnegative,
then so are the $F_{N,k}\,$.
\end{enumerate}

\section{Clebsch-Gordan and 6--$j$ Symbols}
\label{sec:A5}

Let us now compute the Clebsch-Gordan coefficients for the 
decomposition into irreducible representations of the 
{\em symmetric} product of two generic representations.
In general we can write
\be
\hyper{k}{(\beta)_k}{(\alpha)_{k}}(\bz) \;
\hyper{l}{(\delta)_l}{(\gamma)_{l}}(\bz) \;
\, = \; \, \sum_{m}
\,\; \Clebsch{k,l,m}{(\beta)_k;(\delta)_l;(\epsilon)_m}%
{(\alpha)_k;(\gamma)_l;(\zeta)_m}
 \, \hyper{m}{(\zeta)_m}{(\epsilon)_m}(\bz)\; .
\label{CYYY}
\ee
Using the orthogonality relations~(\ref{eq:ortho}) we obtain
\be
\Clebsch{k,l,m}{(\beta)_k;(\delta)_l;(\epsilon)_m}%
{(\alpha)_k;(\gamma)_l;(\zeta)_m} \; =\; 
\int d\Omega(\bz,\bzbar) \;\,
     \hyper{k}{(\beta)_k}{(\alpha)_{k}}(\bz) \,
     \hyper{l}{(\delta)_l}{(\gamma)_{l}}(\bz) \,
     \hyper{m}{(\epsilon)_m}{(\zeta)_m}(\bz)
\label{Cintegral}
\; \mbox{.}
\ee
This integral can be computed explicitly. We get (see Appendix \ref{app:CG})
\begin{eqnarray}
&& \hskip -1truecm \Clebsch{k,l,m}{(\beta)_k;(\delta)_l;(\epsilon)_m}%
{(\alpha)_k;(\gamma)_l;(\zeta)_m}  \; =  \;
     {(N-1)!\, \mu_{N,k} \, \mu_{N,l} \, \mu_{N,m}\, k!\, l!\, m! \over 
      (N-1 + k + l + m)!} 
\nonumber \\[2mm]
&& \hskip -0.5truecm
\times \; \sum_{i = i_{min}}^{i_{max}} 
{k \choose h} {l \choose i} {m \choose j} \,
I_{N,k,(\beta)_k;(\rho)_h (\sigma)_{k-h}}^{
        \hphantom{N,k,}(\alpha)_k;(\mu)_i(\nu)_{k-i}}\;
I_{N,l,(\delta)_l;(\mu)_i (\eta)_{l-i}}^{
        \hphantom{N,l,}(\gamma)_l;(\tau)_j(\sigma)_{k-h}}\;
I_{N,m,(\epsilon)_m;(\tau)_j(\nu)_{k-i}}^{
        \hphantom{N,m,}(\zeta)_m;(\rho)_h(\eta)_{l-i}}
\label{eq:CG}
\end{eqnarray}
if $\; \left| l - k \right| \leq m \leq l + k \;$;
the Clebsch-Gordan coefficient
vanishes otherwise. In Eq.\  \reff{eq:CG} $h=m-l+i$,
$j=m-k+i$, $i_{max} = \min(k,l,l+k-m)$ and 
$i_{min} = \max(0,l-m,k-m)$.

In the following we will be interested in the scalar quantity \be
{\cal C}^2_{N; \; k,l,m} \; =\; {\cal C}_{N;  \; k,l,m} \cdot
{\cal C}_{N;  \; k,l,m} 
\; \mbox{.}
\label{Csquared}
\ee
The general formula is reported in 
Appendix \ref{app:CG} [see~(\ref{eq:C2general})].
A particularly simple case is $m = l + k$:
\be
{\cal C}_{N; \; k,l,l+k}^{2} \; = \; {\cal N}_{N,l+k}\; \frac{ \mu_{N,k}^{2} \;
           \, \mu_{N,l}^{2} }{ \mu_{N,l+k}^{2} }
\label{eq:C2kllk}
\; \mbox{,}
\ee
which can be
obtained directly from~(\ref{eq:CG}), using~(\ref{eq:trace}).
If $k = 1$ this gives
\be
{\cal C}_{N; \; 1,l,l+1}^{2} \; = \; {N(N^2-1)\over (N+2l)}
   {N + l - 1\choose l}^2
\; \mbox{.}
\label{eq:C21ll1}
\ee
Notice that for $N=2$, because of the symmetry \reff{simmetriaN=2},
the Clebsch-Gordan coefficients 
$\Clebsch{k,l,m}{(\beta)_k;(\delta)_l;(\epsilon)_m}%
{(\alpha)_k;(\gamma)_l;(\zeta)_m}$ --- and 
therefore also the scalar invariants ${\cal C}_{N; \; k,l,m}^{2}$ --- 
vanish if $k+l+m$ is odd. 

Let us now derive two important properties of the 
Clebsch-Gordan coefficients. 
Using their definition in terms of 
hyperspherical harmonics and the completeness relation
\reff{complrel}
we can easily prove the crossing relation 
\be
 \sum_{p=0}^{\infty} \Clebsch{p,k,l}{(\delta)_p;(\epsilon)_k;(\zeta)_l}{
                        \!(\alpha)_p;(\beta)_k;(\gamma)_l} \,
        \Clebsch{p,m,n}{(\alpha)_p;(\sigma)_m;(\tau)_n}{
                        (\delta)_p;(\mu)_m;(\nu)_n} \,
 =\, \sum_{p=0}^{\infty} 
        \Clebsch{p,k,m}{(\delta)_p;(\epsilon)_k;(\sigma)_m}{
                        \!(\alpha)_p;(\beta)_k;(\mu)_m} \,
        \Clebsch{p,l,n}{(\alpha)_p;(\zeta)_l;(\tau)_n}{
                        (\delta)_p;(\gamma)_l;(\nu)_n}
  \;.
\label{Ccrossing}
\ee
A second relation, which follows immediately from Schur's lemma,
is
\be
\Clebsch{k,l,m}{(\gamma)_k;(\delta)_l;(\zeta)_m}{
                (\alpha)_k;(\beta)_l;(\epsilon)_m} \;
\Clebsch{k,l,n}{(\alpha)_k;(\beta)_l;(\nu)_n}{
                (\gamma)_k;(\delta)_l;(\mu)_n} \; =\; 
           \delta_{mn}\, {1\over {\cal N}_{N,m}}\,
           I_{N,m,(\zeta)_m;(\nu)_m}^{
              \hphantom{N,m}(\epsilon)_m;(\mu)_m} \;
       {\cal C}_{N;k,l,m}^2
  \;.
\label{CSchur}
\ee
Finally, using the completeness relation \reff{complrel} and 
Eqs.\  \reff{Cintegral}, \reff{YdotY}, 
it is easy to verify the identity
\be
\sum_{k = 0}^{\infty} \; 
  {\cal C}^2_{N; \; k,l,m} \; =\; {\cal N}_{N,l}\; {\cal N}_{N,m}
\label{sumkCquadro}
\; \mbox{.}
\ee

Let us now introduce the 
6--$j$ symbols (also called Racah symbols). They are $U(N)$-scalars defined 
by\footnote{In the terminology used in works dealing with 
high-temperature expansions \cite{Domb_DG}, the 6--$j$ symbols 
are the group factors associated to the so-called $\alpha$-topology.}
\begin{eqnarray}
{\cal R}_N (a,b,c;d,e,f) &=& 
   \Clebsch{a,d,c}{(\eta)_a;(\lambda)_d;(\mu)_c}{
                   (\alpha)_a;(\beta)_d;(\gamma)_c} \;
   \Clebsch{a,b,e}{(\alpha)_a;(\nu)_b;(\rho)_e}{
                   (\eta)_a;(\delta)_b;(\epsilon)_e} \;
  \nonumber \\[2mm]
   && \quad 
   \times\, \Clebsch{d,e,f}{(\beta)_d;(\epsilon)_e;(\sigma)_f}{
                   (\lambda)_d;(\rho)_e;(\zeta)_f} \;
   \Clebsch{b,c,f}{(\delta)_b;(\gamma)_c;(\zeta)_f}{
                   (\nu)_b;(\mu)_c;(\sigma)_f}
   \;.
\label{Racahdef}
\end{eqnarray}
The tetrahedral symmetry, which is enjoyed by the standard 6--$j$ symbols 
\cite{Edmonds}, is trivially true also for  our definition.

We have not been able to compute a general formula for the 6--$j$ symbols,
but we have computed a very large class of special cases.
This class is sufficient
for computing the high-temperature expansion
of the $U(N)$-invariant $\sigma$-models, in general dimension $d$,
up to rather high order. 

We begin by deriving a completeness relation for the 6--$j$ symbols. 
Using Eqs.\  \reff{Ccrossing} and \reff{CSchur}, and the definition
\reff{Racahdef} we get
\be
\sum_a {\cal R}_N (a,b,c;d,e,f)\, =\, {1\over {\cal N}_{N,f}} \,
       {\cal C}^2_{N;d,e,f} \, {\cal C}^2_{N;b,c,f}
  \;.
\label{completeR}
\ee
A general formula for ${\cal R}_N (l+p,l,p;r,m,k)$ for 
arbitrary $l,p,r,m,k$ is reported in App.\ A.4 [see Eq.\  \reff{Rrisultato}]. 
It is particularly simple to derive all 6--$j$ symbols in which one of the 
spins is 1. In this case we need to compute the coefficients
${\cal R}_N(1,l+a,k+b;k,l,m)$ with $a,b=\pm 1,0$. 
Using the tetrahedral symmetry one can immediately recognize that
all cases can be computed using Eq.\  \reff{Rrisultato} of App.\ A.4, 
except when
$a=b=0$. In this case, however, we can use the completeness relation 
\reff{completeR} to write
\begin{eqnarray}
{\cal R}_N(1,l,k;k,l,m) &=& 
   {1\over {\cal N}_{N,k}} {\cal C}^2_{N;k,l,m}
   {\cal C}^2_{N;1,k,k}  \nonumber \\[2mm]
&& - 
   {\cal R}_N(1,l-1,k;k,l,m) - 
   {\cal R}_N(1,l+1,k;k,l,m) \;. \quad
\end{eqnarray}

\section{Applications}

The formalism we have presented in the previous sections can be applied to the
generation of high-temperature series for 
$U(N)$-invariant spin models defined on $CP^{N-1}$. 
The idea is to expand the Gibbs weight $e^{-\beta H}$ for the given 
spin model in terms of hyperspherical harmonics, and then use their
orthogonality to reduce the number of non-vanishing terms. 
We will 
consider here the Hamiltonians that are commonly used in the study
of these models. 

The ``quartic" Hamiltonian is given by
\be
H = - \sum_{\<xy\>} |\bzbar_x\cdot\bz_y|^2\; ,
\label{Hstandard}
\ee
where $\bzbar_x\cdot\bz_x= 1$ and the sum is over all lattice links $\<xy\>$. 
The Hamiltonian \reff{Hstandard} is invariant under global 
$U(N)$ transformations and under the local $U(1)$-gauge transformations
defined by
\be 
\bz_x\to e^{i\alpha_x}\, \bz_x,
\ee
and therefore defines a theory on $CP^{N-1}$.
Notice that for $N=2$, using Eq.\ \reff{CP1S2}, we have 
$|\bzbar_x\cdot\bz_y|^2 = (1 + s_x\cdot s_y)/2\,$. Thus 
the model with Hamiltonian \reff{Hstandard} and $N=2$ is equivalent to the 
$N$-vector model with $N=3$.

A second possibility is the so-called ``gauge" Hamiltonian. In this case
one associates to each lattice link a real gauge field 
$\theta_{xy}\in[0,2\pi[$ and 
considers 
\be
H = - \mbox{\rm Re}\, \sum_{\<xy\>} (\bzbar_x\cdot\bz_y) \, e^{i\theta_{xy}},
\label{Hgauge}
\ee
where $\bzbar_x\cdot\bz_x= 1$ and the sum is over all lattice links $\<xy\>$. 
The Hamiltonian \reff{Hgauge} is invariant under the local 
transformations
\be
\bz_x\to e^{i\alpha_x}\, \bz_x, \qquad 
 \theta_{xy} \to \theta_{xy} + \alpha_x - \alpha_y,
\ee
and therefore defines a theory on $CP^{N-1}$.
If one considers only correlations of the $\bz$-field, one can integrate
out the gauge field $\theta_{xy}$ obtaining an effective 
Hamiltonian
\be
H = - {1\over \beta} \sum_{\<xy\>}
  \log\left[ I_0\left(\beta |\bzbar_x\cdot\bz_y|\right)\right].
\ee
For the Hamiltonians \reff{Hstandard} and \reff{Hgauge},
using \reff{Fgeneralformula}--\reff{FTaylor}, it is 
possible to compute the expansion coefficients $F_{N,k}$.
We need to compute $F_{N,k}$ for the functions
$\exp \left[\beta \left|\bzbar \cdot \bw\right|^2 \right] $ and
$I_0 \left( \beta \left|\bzbar \cdot \bw \right| \right)$.
In the first case  we obtain
\be
F_{N,k} \; = \,\; 
{(N-1)! \, k!\over (N-1+2k)!}\, \beta^k \,
{}_1F_1(k+1;N+2k;\beta)\;,
\label{eq:Fstandard}
\ee
where $\hphantom{}_{1}F_{1}$ is the {\em confluent}\/ ({\em degenerate}\/)
{\em  hypergeometric
function}.\footnote{See \cite{GR}, pp.\ 1058--1059.}
In the second case
the integration gives
\be
F_{N,k} \; = \; 
(N-1)!\, \left({\beta\over2}\right)^{1-N}
    I_{N+2k-1}(\beta)\;,
\label{eq:Fgauge}
\ee
where $I_{\nu}$ is the {\em modified Bessel function}.

\appendix 
\section{Properties of Hyperspherical Harmonics}
\label{app.A}

\subsection{Some basic formulae}
\label{app:HH}

Let us begin by computing the dimension of the linear space $E_{N,k}$.
This can be done by computing the dimension of the space
of {\em all}\/ completely symmetric tensors of rank $k$ in two sets of 
indices, and then
subtracting from it the number of independent trace conditions that have
to be imposed in order to ensure the tracelessness of these tensors. 
The number of linearly independent symmetric tensors is given by 
$\,{N+k-1 \choose{k}}^2\,$ 
and the number of traces is
given by $\,{N+k-2 \choose{k-1}}^2\,$.
Therefore we obtain
\begin{eqnarray}
{\cal N}_{N,k}  \,\equiv\,
\dim E_{N,k} &=& {N+k-1 \choose{k}}^2 \,-\, {N+k-2 \choose{k-1}}^2
                                                       \\[3mm]
             &=& (N-1) (N-1+2k) 
     \left[ (N-2+k)!\over k! (N-1)!\right]^2.
                                                      \label{eq:A.1}
\end{eqnarray} 
This proves Formula \reff{eq:dim}. For $N=2$ we have 
${\cal N}_{2,k} = 2 k + 1$, as expected on the basis of the 
isomorphism $S^2\simeq CP^1$.

Let us now compute the integral of monomials in
$\bz$ and $\bzbar$, i.e.\  of 
\be
\int d\Omega(\bz,\bzbar)\;
   z^{\alpha_1}\ldots z^{\alpha_k} 
   \zbar_{\beta_1}\ldots \zbar_{\beta_l}\;.
\ee
It is trivial to see that the integral is zero if $k\not=l$. 
To compute its value for $k=l$, we introduce an 
arbitrary complex vector $A^\alpha$, and define
\be
I_k(A) = \int d\Omega(\bz,\bzbar)\;
 (\overline{A}\cdot \bz)^k \, (\bzbar\cdot A)^k\;.
\label{IkA}
\ee
Since the measure $d\Omega(\bz,\bzbar)$ is $U(N)$-invariant, 
we have $I_k(A) = I_k(UA)$ for any $U\in U(N)$, so that $I_k(A)$
depends only on $\overline{A}\cdot A$. Moreover,
$I_k(A)$ is manifestly a homogeneous function of degree $k$
in $A$ and $\overline{A}$. Hence we must have 
$I_k(A) = J_k (\overline{A}\cdot A)^k$ for some constant
$J_k$. Now, since $\bzbar\cdot\bz = 1$, we have
\be
{\partial\over \partial \overline{A}_\alpha} 
{\partial\over \partial A^\alpha}  I_k\left(A\right) \; =\; \,
k^2  I_{k-1} \left(A\right)\; .
\ee
A recursion relation for $J_k$ immediately follows:
\be
J_k \;=\; { k\over N + k - 1} \, J_{k-1}\;.
\ee
Using $J_0=1$ we obtain the general solution
\be
J_k \; = \; {k! (N-1)!\over (N-1+k)!}\;.
\ee
Taking then $k$ derivatives with respect to $A$ and 
with respect to $\overline{A}$ in Eq.\ \reff{IkA} we obtain
the well-known result
\be
\int d\Omega(\bz,\bzbar)
   z^{\alpha_1}\ldots z^{\alpha_k} 
   \zbar_{\beta_1}\ldots \zbar_{\beta_k} \; =
	\; {(N-1)!\over (N-1+k)!}
        \; \left( \delta^{\alpha_{1}}_{\beta_1} \cdots
        \delta^{\alpha_k}_{\beta_k} +  \; \ldots \; \right)
\label{eq:sigma}
\ee
where the terms in parentheses correspond to the $k!$
different pairings of the indices.

Let us now prove the orthogonality relation \reff{eq:ortho}. This 
is completely equivalent to proving that
for arbitrary completely symmetric (in two sets of indices)
and traceless tensors $T_{N,k}$ and $U_{N,l}$ we have
\be
\int d\Omega(\bz,\bzbar) \; \;
        (Y_{N,k} \cdot T_{N,k} )
        (Y_{N,l} \cdot U_{N,l} )\; =\; 
        \delta_{k l}\;
          T_{N,k} \cdot U_{N,k}
\label{eq:2Y}
\; \mbox{.}
\ee
To see this, let us first use the definition~(\ref{eq:Ydef}) 
and let us notice that the ``Traces'' terms  do not give any 
contribution due to the tracelessness of $T_{N,k}$ and $U_{N,l}$. Thus the 
l.h.s.\  in Eq.\ \reff{eq:2Y} becomes simply 
\begin{eqnarray}
& & \mu_{N,k}\;\mu_{N,l}\; \left[ \int d\Omega(\bz,\bzbar) \; \,
z^{\alpha_1}\ldots z^{\alpha_k} 
z^{\beta_1}\ldots z^{\beta_l }
\zbar_{\gamma_1} \ldots \zbar_{\gamma_k}
\zbar_{\delta_1} \ldots \zbar_{\delta_l}
\right]
\;
T_{N,k,(\alpha)_k}^{ \hphantom{N,k}(\gamma)_k}\; 
U_{N,l,(\beta)_l}^{ \hphantom{N,l}(\delta)_l}
\; \mbox{.} \nonumber \\[2mm] & &
\end{eqnarray}
We can then use Eq.\ \reff{eq:sigma}. The only non-vanishing contributions
come from those terms which do not contain $\delta^{\alpha_i}_{\gamma_j}$
or $\delta^{\beta_i}_{\delta_j}$; such terms exist only if $l=k$. 
In this last case there are $(k!)^2$ equivalent contractions and we 
end up with
\be
\delta_{kl} \;\, \mu_{N,k}^2 \; {(N-1)! (k!)^2\over (N-1 + 2k)!}
 \; T_{N,k} \cdot U_{N,k} \; = \;
\delta_{kl} \; T_{N,k} \cdot U_{N,k}
\; \mbox{.}
\ee
Using the expression (\ref{eq:mu}) for the normalization factor
$\mu_{N,k}$, we obtain the orthogonality relation~(\ref{eq:ortho}) for
the $Y$'s.

The general formula for the hyperspherical harmonics
\cite{Rossi} can be obtained by using
the fact that they are completely symmetric and 
traceless. The complete symmetry, together with the needed transformation
properties under $U(N)$, implies an expansion of the form
\be
\hyper{k}{(\beta)_k}{(\alpha)_k} (\bz) 
\; =\;
\mu_{N,k}\; \sum_{s=0}^{k}\; 
A_{N,k;s} \, P_{(k;s)(\beta)_k}^{\hphantom{(k;s)}(\alpha)_k} \left(\bz\right)
\; \mbox{,}
\label{Ygeneral}
\ee
where
\be
P_{(k;s)(\beta)_k}^{\hphantom{(k;s)}(\alpha)_k} \left(\bz\right)
\equiv\;
\delta^{\alpha_1}_{\beta_1}\ldots
\delta^{\alpha_s}_{\beta_s} z^{\alpha_{s+1}} \ldots z^{\alpha_k}
\zbar_{\beta_{s+1}}\ldots \zbar_{\beta_k}
 +\, \hbox{permutations}
\label{Pksdefinition}
\ee
and the number of permutations necessary to make $P_{(k;s)}$ completely
symmetric is $s!\; {k\choose{s}}^2$.
Now we impose the tracelessness.
We first note that
\be
P_{(k;s)\alpha (\beta)_{k-1}}^{\hphantom{(k;s)}\alpha (\gamma)_{k-1}} 
  \left(\bz\right)  \; =\;
P_{(k-1;s)(\beta)_{k-1}}^{\hphantom{(k-1;s)} (\gamma)_{k-1}} 
  \left(\bz\right)  \; +\;
(N + 2k - s - 1)
P_{(k-1;s-1)(\beta)_{k-1}}^{\hphantom{(k-1;s-1)} (\gamma)_{k-1}} 
  \left(\bz\right)  
\; \mbox{,}
\ee
with the understanding that $P_{(k;s)} = 0$ for $s > k$ and $s<0$. From
the tracelessness of the hyperspherical harmonics we obtain the
recursion relation
\be
A_{N,k;s} \,+\, (N + 2k - s -2) \, A_{N,k;s+1} \;=\;0
\; \mbox{.}
\label{Arecur}
\ee
Imposing the normalization $\,A_{N,k;0} = 1$, we find
\be
A_{N,k;s} \; = \; (-1)^s \; {(N + 2 k - s - 2)!\over (N+2k - 2)!}
        \; \mbox{.}
\label{eq:defA}
\ee

\vskip 3mm

Let us now discuss the relation between the hyperspherical harmonics and 
the Jacobi polynomials. From Section \ref{sec2}
we know that $\hyper{k}{1\ldots 1}{1\ldots 1}\left(\bz\right)$
is the restriction to the
complex unit sphere of a harmonic polynomial of degree 
$k$ in $\bz$ and $\bzbar$. 
Moreover, it depends only on $|z^1|$ because of the gauge invariance 
$\bz \to e^{i\alpha} \bz$. Therefore it can be written as $r^{2k}
P_k\left(|z_1|^2/r^2\right)$ where
$r = \left|{\bz}\right|$.
Requiring the polynomial to satisfy Laplace's equation 
we get for $P_k(x)$ the equation
\be
x (1 - x)\, {d^2 P_k\over d x^2}\, + (1 - Nx)
\, {d P_k\over dx}\, + 
k\,\left(N+k-1\right)\, P_k\; =\; 0.
\ee
The regular solution of this equation is the Jacobi polynomial 
$P_k^{(N-2,0)}\left(2x-1\right)$ (see \cite[p.\ 1036]{GR}). 
The normalization is fixed by the requirement that
\be
\hyper{k}{1\ldots 1}{1\ldots 1}\left(\bz\right) \; =\; 
 \mu_{N,k} \, |z^1|^{2k} \, + \, \hbox{lower-order terms}\; .
\ee
We thus get (see Formula 8.962.1b in \cite[p.\ 1036]{GR})
\be
\hyper{k}{1\ldots 1}{1\ldots 1}\left(\bz\right) \; =\; 
\mu_{N,k} {k! (N-2 + k)!\over (N-2 + 2k)!} 
\; P_k^{(N-2,0)}\left(2 |z_1|^2 - 1\right)
\ee
which, using the fact that
\be
P_k^{(N-2,0)}\left(1\right) \; = \; {N-2+k\choose{k}}
\label{Jacobiin1}
\ee
gives \reff{YGegen}.

Note that we could have derived \reff{Ygeneral} by
using \reff{YGegen} and the
expansion of the Jacobi polynomials.

\subsection{Expansions in Terms of Hyperspherical Harmonics}
\label{app:Exp}

We want now to discuss the convergence of the expansion \reff{genericfexp}.
We will begin by noticing that, given a generic complex
tensor
$T_{N,k,(\beta)_k}^{\hphantom{N,k,}(\alpha)_k}$, the hyperspherical harmonics 
satisfy the inequality
\be
\left|\, T_{N,k}\cdot Y_{N,k}(\bz)\,\right|^2\, \leq \;
 (T_{N,k}\cdot T_{N,k}^*)\; {\cal N}_{N,k}\; ,
\label{TSchwarz}
\ee
where the {\em adjoint} tensor $T^*_{N,k}$ is given by
\be
\left(T^{*}_{N,k}\right)^{(\alpha)_k}_{(\beta)_k} \; =\;
\overline{\left(T_{N,k}\right)_{(\alpha)_k}^{(\beta)_k}} \; ,
\ee
and the overline indicates complex conjugation.
Eq.\  \reff{TSchwarz} follows immediately from 
Schwarz's inequality and \reff{YdotY}.
Moreover, equality in \reff{TSchwarz} is possible only
for those $\bz$'s for which
\be
T_{N,k} \; =\; \gamma\ Y_{N,k} (\bz)
\ee
for some constant $\gamma$. This requires first of all that
$\,T_{N,k}\,$ be completely symmetric in both sets of indices 
and traceless. The
constant $\gamma$ is easily obtained squaring the previous relation:
\be
|\gamma|^2 \; = \; {\left(T_{N,k}^*\cdot T_{N,k}\right)\over {{\cal N}_{N,k}} }
\; \mbox{.}
\ee

Now let us consider the special case
$\, T_{N,k} = Y_{N,k}(\bx)$ with $\bx 
= \left( 1\mbox{,}\; 0\mbox{,}\; \ldots\mbox{,}
\; 0\right)$
and $k\geq 1$. In this case we can rephrase Eq.\ \reff{TSchwarz} 
in terms of Jacobi polynomials [using Eqs.\ \reff{YdotY}, 
\reff{YGegen} and \reff{Y_dot_Y}] as
\be
|P^{(N-2,0)}_k(t)|^2 \;\leq\; |P^{(N-2,0)}_k(1)|^2\;.
\label{P_Schwarz}
\ee
Equality in this case is possible only if 
\be
Y_{N,k} \left(\bx \right) \; = \; \gamma \,
Y_{N,k} \left(\bz\right)
\; \mbox{,}
\label{equationYsYw}
\ee
with $|\gamma|=1$. We will now prove that, for $N>2$ 
this equation is satisfied only if $\gamma=1$ and $\bz = e^{i\theta} \bx$. 
For $N=2$ there is a second solution:
$\gamma = (- 1)^k$ and $\bz = e^{i\theta} \widetilde{\bx}$, with 
$\widetilde{\bx} = (0,1)$.

For $b\not = 1$ we have from Eq.\ \reff{equationYsYw}
\be 
0 = \hyper{k}{b\ldots b}{1\ldots 1} (\bz) = \mu_{N,k} \, 
(z^1)^k (\zbar_b)^k,
\ee
that implies either $z^1 = 0$ or $z^b = 0$. 
Now, if $z^1 = 0$, from Eqs.\  \reff{equationYsYw} and 
\reff{YGegen}, we obtain 
\be 
P_k^{(N-2,0)}(1) \;=\; \gamma \,
P_k^{(N-2,0)}(-1) \;.
\label{eqA.26}
\ee
Now $P_k^{(N-2,0)}(-1) = (-1)^k$, while $P_k^{(N-2,0)}(1)$ is given
in Eq.\  \reff{Jacobiin1}. For $N>2$ and $|\gamma|=1$, 
one immediately verifies that 
Eq.\  \reff{eqA.26} is never satisfied. Therefore $z^1\not=0$, so that 
$z^b = 0$ for all $b\not = 1$. The result immediately follows. 
For $N=2$, using Eq.\  \reff{simmetriaN=2}, it is easy to verify that 
both $e^{i\theta}\bx$ and $e^{i\theta} \widetilde{\bx}$ 
satisfy Eq.\  \reff{equationYsYw}.

Thus, we have shown that equality in Eq.\ \reff{P_Schwarz} holds only
for the case $t=1$ (respectively $t=\pm 1$) for $N>2$ (respectively
$N=2$). We therefore have [since $P^{(N-2,0)}_k(1)>0$]
\be
|P^{(N-2,0)}_k(t)| <\; P^{(N-2,0)}_k(1)
\label{Cbound}
\ee
for $-1<t<1$ and $k\geq 1$.

To discuss the convergence of the series
\reff{genericfexp} let us first notice that
using the completeness relation \reff{complrel} and 
Eq.\  \reff{tildef}  we get 
\be
\sum_{k = 0}^{\infty}\; 
  \widetilde{f}^{\phantom{k,}(\alpha)_k}_{k,(\beta)_k}  \,
  \widetilde{f}_{k,(\alpha)_k}^{\phantom{k,}(\beta)_k}  \; = \;
       \int d\Omega\left(\bz,\bzbar\right) \; 
        \left| f\left(\bz\right) \right|^2
\; \mbox{,}
\ee
which is the Plancherel identity for harmonic analysis in $CP^{N-1}$.
Now let us consider, instead of $f$, the function ${\cal L}^n f$
where ${\cal L}$ is the Laplace-Beltrami operator. In this case
$\, \widetilde{f}^{\phantom{k,}(\alpha)_k}_{k,(\beta)_k}  \,$
is replaced by 
$\,(\lambda_{N,k})^n \,
\widetilde{f}^{\phantom{k,}(\alpha)_k}_{k,(\beta)_k}  \,$,
and thus we obtain
\be
\sum_{k = 0}^{\infty}\; (\lambda_{N,k})^{2n}\,
  \widetilde{f}^{\phantom{k,}(\alpha)_k}_{k,(\beta)_k}  \,
  \widetilde{f}_{k,(\alpha)_k}^{\phantom{k,}(\beta)_k}  \; = \;
       \int d\Omega\left(\bz,\bzbar\right) \; 
        \left| {\cal L}^nf\left(\bz\right) \right|^2
\; \mbox{.}
\ee
If now $f$ is a $C^\infty$ function, the integral is finite 
for all $n$. Thus the sum on the l.h.s.\  is converging for all $n$. 
Since $\lambda_{N,k} \sim k^2$ for $k\to\infty$, we get that
$ k^{2 n}  \widetilde{f}_k\cdot \widetilde{f}_k  \to 0$ in the same
limit, for every $n$.
This implies that all coefficients $\widetilde{f}_k$
decrease faster than any inverse power of $k$.
To prove the convergence of the series~\reff{genericfexp}
it is then enough to notice 
that $|Y_{N,k} (\bz) |\leq \left({\cal N}_{N,k}
\right)^{1/2}$ 
and that ${\cal N}_{N,k}$ behaves as $\,k^{2N-3}\,$ for large $k$.

Let us now discuss the properties of the coefficients $F_{N,k}$
in \reff{Fgeneralformula}. The second property follows immediately from
the previous discussion. We want now to prove that, if $f\left(t\right)$ is 
positive for $t\in[0,1]$, then $\, | F_{N,k} | < F_{N,0}\, $ 
for $\, k\geq 1$. 
Indeed, from Eqs.\ \reff{Fgeneralformula} and \reff{Cbound} we get
\begin{eqnarray}
|F_{N,k}| &\leq & (N-1) \int_0^{1} dt\;\,
        (1 - t)^{N-2} \; f(\sqrt{t}) \;
\left| {P_k^{(N-2,0)}(t)\over P_k^{(N-2,0)}(1)}\right| \nonumber \\
&<& (N-1) \int_0^{1} dt\;\,
         (1 - t)^{N-2} \; f(\sqrt{t})  =\, F_{N,0}
\; \mbox{.}
\end{eqnarray}

\subsection{Some Useful Contractions}

We are now going to report some useful formulae for the contractions
of various products of hyperspherical harmonics. To simplify the results
we will introduce some additional notation.

Given three irreducible (completely symmetric and traceless) tensors
$T_{N,k}$, $U_{N,l}$, and $V_{N,m}$ --- respectively of rank 
$k$, $l$ and $m$ --- we want
to construct a scalar, i.e.\  a quantity that is invariant under 
$U(N)$ transformations. It is easy to see that there are many
ways of doing this, each one of them characterized by an integer 
$i$. We define
\begin{eqnarray}
&& \hskip -1truecm 
   \left[T_{N,k}\cdot U_{N,l} \cdot V_{N,m}\right]_i \equiv 
\nonumber \\[2mm]
&& \qquad 
   T^{\hphantom{N,k} (\alpha)_i (\beta)_{k-i} }_{
               N,k,(\zeta)_{m-l+i} (\gamma)_{k-i+l-m}   } \;
   U^{\hphantom{N,l} (\gamma)_{k-i+l-m} (\delta)_{m+i-k} }_{
               N,l,(\alpha)_i (\epsilon)_{l-i}  } \;
   V^{\hphantom{N,m} (\epsilon)_{l-i} (\zeta)_{m-l+i} }_{
               N,m,(\beta)_{k-i} (\delta)_{m+i-k}   }.
\label{eqcontraction}
\end{eqnarray}
This contraction is defined for
$0\leq i \leq (k,l) \leq m + i\leq k + l$. In the previous 
expression, $i$ counts the number of upper indices 
of $T_{N,k}$ that are contracted with the lower indices 
of $U_{N,l}$. Once $i$ is given, all other contractions 
are completely fixed.

The definition \reff{eqcontraction} can be generalized by writing 
\begin{eqnarray}
&& \hskip -2truecm 
   \left[T_{N,k+p,(\nu)_p}^{\hphantom{N,k+p,}(\mu)_p}
        \cdot U_{N,l} \cdot V_{N,m}\right]_i \equiv 
\nonumber \\[2mm]
&&  
   T^{\hphantom{N,k+p,} (\mu)_p (\alpha)_i (\beta)_{k-i} }_{
               N,k+p,(\nu)_p (\zeta)_{m-l+i} (\gamma)_{k-i+l-m}   } \;
   U^{\hphantom{N,l,} (\gamma)_{k-i+l-m} (\delta)_{m+i-k} }_{
               N,l,(\alpha)_i (\epsilon)_{l-i}  } \;
   V^{\hphantom{N,m,} (\epsilon)_{l-i} (\zeta)_{m-l+i} }_{
               N,m,(\beta)_{k-i} (\delta)_{m+i-k}   }.
\end{eqnarray}
We also introduce a second useful notation:
\be
\ZZ{k}{(\beta)_k}{(\alpha)_k}(\bz) \equiv 
  \; z^{\alpha_1}\ldots z^{\alpha_k}\zbar_{\beta_1} \ldots
     \zbar_{\beta_k}.
\ee

Let us now compute various contractions that will be used in the 
determination of the Clebsch-Gordan and 6--$j$ coefficients.
\begin{itemize}
\item[(i)]
  $B_1(l,m;h,k)$.
\end{itemize}
Let us define
\be
B_1(l,m;h,k) = \left [Y_{n,l}(\bz) \cdot Y_{N,m}(\bz) \cdot Z_{N,l+m-k-h}(\bz)
               \right]_h\;,
\ee
for $(l,m) \geq (h,k)$.
By $U(N)$-invariance it is immediate to see that $B_1$ does not 
depend on $\bz$. Using the property\footnote{This property is easily
obtained recursively, by writing
\be
z^\alpha \zbar_\beta \;
  \hyper{m}{\alpha(\delta)_{m-1}}{\beta(\gamma)_{m-1}} (\bz) \; = \;
  x_m \;
  \hyper{m-1}{(\delta)_{m-1}}{(\gamma)_{m-1}} (\bz) \;,
\ee
and noticing that 
\be
Z_{N,m}(\bz)\cdot Y_{N,m}(\bz)\;=\;\frac{{\cal N}_{N,m}}{\mu_{N,m}}\;=\;
\prod_{k=1}^m \, x_k\;.
\ee
}
\be
z^\alpha \zbar_\beta \;
  \hyper{m}{\alpha(\delta)_{m-1}}{\beta(\gamma)_{m-1}} (\bz) \; = \;
  {{\cal N}_{N,m}\, \mu_{N,m-1}\over {\cal N}_{N,m-1}\, \mu_{N,m}} \;
  \hyper{m-1}{(\delta)_{m-1}}{(\gamma)_{m-1}} (\bz) ,
\label{zzbarY}
\ee
we obtain for $h>k$
\be
B_1(l,m;h,k) = 
  {{\cal N}_{N,m}\ {\cal N}_{N,l}\ \mu_{N,h}^2 \over 
   {\cal N}_h^2\, \mu_{N,m}\ \mu_{N,l}} \;
 \left[ Y_{N,h}(\bz) \cdot Y_{N,h}(\bz) \cdot Z_{N,h-k}(\bz)\right]_h\; .
\ee
Now we expand the first $Y_{N,h}(\bz)$ using Eq.\  \reff{Ygeneral}, and
then use repeatedly Eq.\ \reff{zzbarY}. We obtain finally $(h\ge k)$
\begin{eqnarray}
B_1(l,m;h,k) & = &
  {{\cal N}_{N,m}\ {\cal N}_{N,l}\, \mu_{N,h}^2 \over 
   {\cal N}_h\, \mu_{N,m}\ \mu_{N,l}}\; \sum_{s=0}^{h-k}
   {h\choose s} {h-k\choose s}\ s!\, A_{N,h;s}  
\nonumber \\[2mm]
& =& {{\cal N}_{N,m}\ {\cal N}_{N,l} \over \mu_{N,m}\ \mu_{N,l}}\;
    {(N-2)!\, (N-2+h+k)!\over (N-2+h)!\, (N-2+k)!}\;.
\label{B1risultato}
\end{eqnarray}
For $h < k$ we can use the symmetry
\be
B_1(l,m;h,k) = B_1(m,l;k,h)\;.
\ee
Since the final expression, Eq.\  \reff{B1risultato}, is symmetric in 
$m$ and $l$ and in $h$ and $k$, it is valid also for $h < k$.

\begin{itemize}
\item[(ii)]
$B_2(s;i,l,m,n)$.
\end{itemize}
Let us define
\be
B_2(s;i,l,m,n) =\; \left[ P_{(l;s)}(\bz)\cdot Y_{N,m}(\bz) \cdot
                          Y_{N,n}(\bz)\right]_i\;,
\ee
for $0\le i \le (m,l) \le n+i \le m + l$ and $0 \le s \le l$.

The computation is simple. Indeed, using the definition \reff{Pksdefinition}
of $\,P_{(l;s)}(\bz)$ we can easily express $B_2$ in terms of $B_1$. If
$s\leq l - |m-n|$ we obtain
\begin{eqnarray}
\hskip -1truecm
B_2(s;i,l,m,n) & = &
 \sum_{r=r_{min}}^{r_{max}} {i\choose r} {n-m+i\choose r}
         {l - i\choose s - r} {m - n + l - i\choose s - r}
 \nonumber \\ [2mm]
 && \times\, r!\, (s-r)!\, B_1(m,n;n-l+i+s-r,r+m-i)\;,
\end{eqnarray}
where $r_{min} = \max(0,i+s-l,i+s-l+n-m)$,
      $r_{max} = \min(s,i,i+n-m)$. Otherwise (if $s > l - |m-n|$)
we get
\be
B_2(s;i,l,m,n) = 0\;.
\ee

\begin{itemize}
\item[(iii)]
$B_3(i;l,m,n)$.
\end{itemize}
Let us define
\be
B_3(i;l,m,n) =\; \left[ Y_{N,l}(\bz) \cdot Y_{N,m}(\bz) \cdot Y_{N,n}(\bz)
                 \right]_i\;,
\label{B3_def}
\ee
for $0\le i \le (m,l)\le n+i \le m + l$.

Using Eq.\  \reff{Ygeneral} we rewrite
$B_3$ in terms of $B_2$. It is easy to see that 
\be
B_3(i;l,m,n) = 0
\ee
if $l < |m-n|$ or $l > m + n$; otherwise we obtain
\be
B_3(i;l,m,n) = \mu_{N,l}\, \sum_{s=0}^{l-|m-n|} 
       A_{N,l;s}\, B_2(s;i,l,m,n).
\label{B3risultato}
\ee
Notice the following symmetries of $B_3$:
\be
B_3(i;l,m,n) =\; B_3(l-i;l,n,m) =\; B_3(n+i-l;m,n,l) =\;
                 B_3(m+l-n-i;m,l,n).
\ee
These symmetries are very useful to check the correctness of our result
\reff{B3risultato} since our method of computation gives an expression
that does not have any of these symmetries in an obvious way.

\begin{itemize}
\item[(iv)]
$B_4(i,j;p,r,m)$.
\end{itemize}
Let us define
\begin{eqnarray}
\hskip -1truecm
B_4(i,j;p,r,m) &=& 
  \int d\Omega(\bz,\bzbar)\, d\Omega(\bw,\bwbar) 
\nonumber \\ [2mm]
&& \hskip -1.5truecm 
   \left[Y_{N,p}(\bw)\cdot Z_{N,r-p+i+j}(\bz) \cdot 
               Y_{N,r}(\bw)\right]_i\; 
              \left( Y_{N,m}(\bz) \cdot Y_{N,m}(\bw)\right),
\label{B4_def}
\end{eqnarray}
for $0\le (i,j) \le p \le [(r+i),(r+j)]$.

Using the tracelessness of $Y_{N,m}(\bw)$ and Eq.\ \reff{eq:Ydef}
we can write
\begin{eqnarray}
B_4(i,j;p,r,m) &=& 
  \mu_{N,m}  \int d\Omega(\bz,\bzbar)\, d\Omega(\bw,\bwbar) 
\nonumber \\
&& \hskip -1.5truecm 
      \left[Y_{N,p}(\bw)\cdot Z_{N,r-p+i+j}(\bz) \cdot 
               Y_{N,r}(\bw)\right]_i\; 
              \left( Z_{N,m}(\bz) \cdot Y_{N,m}(\bw)\right).
\end{eqnarray}
We can now perform the integral over $\bz$ using Eq.\  \reff{eq:sigma}.
It is easy to see that $B_4(i,j;p,r,m)$ is non-vanishing 
only if $|r-p| \le m \le r - p + i + j$. If this condition is satisfied
we obtain (the integral over $\bw$ can be dropped)
\begin{eqnarray}
B_4(i,j;p,r,m) &= &
   {\mu_{N,m}\ (N-1)!\, i!\, j!\, (m!)^2\, (r-p+j)!\, (r-p+i)!\over 
      (N-1+r-p+i+j+m)!} 
   \nonumber \\
&& \hskip -1.5truecm
   \sum_{a=a_{min}}^{a_{max}} 
      {1\over a!\, b!\, c!\, d!\, e!\, f!} 
      \left[ Y_{N,p}(\bw) \cdot Y_{N,r}(\bw) \cdot Y_{N,m}(\bw)
      \right]_{r+j-m-a},
\end{eqnarray}
where $b=j-a$, $c=r-p-m-a+i+j$, $d=a+m-j$, $e=p+m+a-r-j$,
      $f=r+j-p-a$, $a_{min} = \max(0,j-m,r-p+j-m)$,
      $a_{max}=\min(j,r-p+j,r-p+i+j-m)$. 
We can now obtain the final result by using Eqs.\ \reff{B3_def}
and \reff{B3risultato}.

We note that, starting from the definition \reff{B4_def} and
using the completeness relation \reff{complrel} and 
Eq.\  \reff{B1risultato}, we obtain the sum rule
\begin{eqnarray}
&& \sum_{m=|r-p|}^{r-p+i+j} B_4(i,j;p,r,m)\; =\; B_1(p,r;p-i,p-j)
\nonumber \\
&& \qquad
 =\; { {\cal N}_{N,r}\, {\cal N}_{N,p} \over 
         \mu_{N,r} \mu_{N,p} }\, 
          {(N-2)!\, (N-2 +2 p - i - j)!\over 
           (N-2+p-i)!\, (N-2+p-j)!}\;. 
\end{eqnarray}

\begin{itemize}
\item[(v)]
$B_5(i,l;p,k,r,m)$.
\end{itemize}
Let us define
\begin{eqnarray}
B_5(i,l;p,k,r,m) &=& 
  \int d\Omega(\bz,\bzbar)\, d\Omega(\bw,\bwbar) 
         \hyper{l}{(\alpha)_l}{(\beta)_l} (\bz)
\nonumber \\
&& \hskip -2.0truecm 
\times \left[Y_{N,l+p,(\beta)_l}^{\hphantom{N,l+p}(\alpha)_l}
         (\bw)\cdot Y_{N,k}(\bz) \cdot 
               Y_{N,r}(\bw)\right]_i\; 
              \left( Y_{N,m}(\bz) \cdot Y_{N,m}(\bw)\right),
\label{B5def}
\end{eqnarray}
for $0\le i\le (p,k) \le r +i \le k + p$.

We expand $Y_{N,k}(\bz)$
using Eq.\  \reff{Ygeneral} and notice that 
$Y_{N,l}(\bz)$ can be replaced by $\mu_{N,l} Z_{N,l}(\bz)$. 
In this way $B_5$ is expressed in terms of $B_4$. 
We obtain finally
\begin{eqnarray}
&& \hskip -1truecm
   B_5(i,l;p,k,r,m) \; = \; \mu_{N,k}\, \mu_{N,l} \,
   \sum_{s=0}^{k-|r-p|} \sum_{a=a_{min}}^{a_{max}} 
   {r-p+i\choose a} {i\choose a} {k-r+p-i\choose s-a} {k-i\choose s-a}
\nonumber \\ [3mm]
&& \quad \times\ (s-a)!\, a!\, A_{N,k;s}\,
   B_4(i-a+l, k-r+p-i-s+a+l; p+l,r,m)
\end{eqnarray}
where $a_{min} = \max(0,s-k+i,s-k+i+r-p)$ and 
      $a_{max} = \min(i,s,r-p+i)$.

Using the fact that $B_4(i,j;p,r,m)$ vanishes unless 
$|r-p| \le m \le r - p + i + j$, we see that
$B_5(i,l;p,k,r,m) = 0$ if $m > k + l - s$ or 
$m < |r - p - l|$.

\subsection{Clebsch-Gordan Coefficients and 6--$j$ Symbols}
\label{app:CG}
 
Let us now discuss the computation of the Clebsch-Gordan coefficients
\reff{eq:CG}. For arbitrary completely symmetric and traceless
tensors $T_{N,k}$, $U_{N,l}$ and $V_{N,m}$, we want to compute
\be
\int d\Omega(\bz,\bzbar)\;\,\left( \,T_{N,k} \cdot Y_{N,k}\left(\bz\right)
       \,\right)\;\left(\,
U_{N,l} \cdot Y_{N,l}\left(\bz\right)\,\right)\;\left(\,
  V_{N,m} \cdot Y_{N,m}\left(\bz\right)\,\right)
\; \mbox{.}
\ee
Using the definition \reff{eq:Ydef} 
of the hyperspherical harmonics this reduces to 
\be
\mu_{N,k} \, \mu_{N,l} \, \mu_{N,m} 
\int d\Omega(\bz,\bzbar)\;\,\left( \,T_{N,k} \cdot Z_{N,k}\left(\bz\right)
       \,\right)\;\left(\,
U_{N,l} \cdot Z_{N,l}\left(\bz\right)\,\right)\;\left(\,
  V_{N,m} \cdot Z_{N,m}\left(\bz\right)\,\right)
\; \mbox{.}
\ee
We can now use Eq.\ \reff{eq:sigma}.
For $|l-m|\le k \le l+m$ the integral becomes
\be
\frac{ (N-1)!\, \mu_{N,k} \, \mu_{N,l} \, \mu_{N,m} }{ 
     (N-1+k+l+m)! } \; \sum_{i=i_{min}}^{i_{max}}
{k\choose h} {l\choose i} {m\choose j} k!\, l!\, m! \,
\left[ T_{N,k} \cdot U_{N,l} \cdot V_{N,m}\right]_i\; ,
\label{integraleCG}
\ee
where $h=m-l+i$, $j = m-k+i$, $i_{max} = \min(k,l,l+k-m)$,
$i_{min} = \max(0,l-m,k-m)$ and we use the notation
introduced in Eq.\ \reff{eqcontraction}. If $k < |l-m|$ or $k > l+m$ 
the integral vanishes.
Formula \reff{eq:CG} immediately follows.

We now discuss the computation of ${\cal C}^2_{N;\; k,l,m}$
for $|l-m|\le k\le l+m$.
Using \reff{Cintegral} we rewrite
\be
{\cal C}^2_{N;\; k,l,m} \;=\; \int d\Omega\left(\bz,\bzbar\right) \; \;
\Clebsch{k,l,m}{(\beta)_k;(\delta)_l;(\zeta)_m}{
                (\alpha)_k;(\gamma)_l;(\epsilon)_m}\,
\hyper{k}{(\alpha)_k}{(\beta)_k}(\bz)\, 
\hyper{l}{(\gamma)_l}{(\delta)_l}(\bz)\, 
\hyper{m}{(\epsilon)_m}{(\zeta)_m}(\bz)\;.
\ee
Using \reff{integraleCG} we obtain 
\begin{eqnarray}
\hskip -1truecm
{\cal C}^2_{N;\; k,l,m} &=& 
\frac{ (N-1)!\, \mu_{N,k} \;\, \mu_{N,l} \;\, \mu_{N,m} }{ 
     (N-1+k+l+m)! } 
\nonumber \\
  &&  \sum_{i=i_{min}}^{i_{max}}
{k\choose h} {l\choose i} {m\choose j} k!\, l!\, m! \,
\left[ Y_{N,k} (\bz) \cdot Y_{N,l} (\bz) \cdot Y_{N,m} (\bz) \right]_i\; ,
\label{eq:C2general}
\end{eqnarray}
where $h,j,i_{min},i_{max}$ are defined as in Eq.\ \reff{integraleCG}.
The final result can be expressed in terms of $B_3(i;k,l,m)$ computed 
in Eq.\ \reff{B3risultato}.

We now describe another way
of computing ${\cal C}^2_{N;\; k,l,m}$. Using
\reff{Cintegral} we can write
\begin{eqnarray}
{\cal C}^2_{N;\; k,l,m} & = & \int d\Omega(\bz,\bzbar) 
d\Omega(\bw,\bwbar) 
\nonumber \\ [1mm]
&& 
\left(Y_{N,k}(\bz)\cdot Y_{N,k}(\bw)\right)
\left(Y_{N,l}(\bz)\cdot Y_{N,l}(\bw)\right)
\left(Y_{N,m}(\bz)\cdot Y_{N,m}(\bw)\right)
\;.
\end{eqnarray}
The integrand is only a function of $|\bwbar\cdot\bz|$. Thus,
using the $U(N)$- invariance of the measure, we can fix
one of the two spins to an arbitrary value. Let us set 
$\bw={\bx}
\equiv \left( 1\mbox{,}\; 0\mbox{,}\; \ldots\mbox{,}
\; 0 \right)$. We obtain, after integrating in $d\Omega(\bw)$,
\be
{\cal C}^2_{N;\; k,l,m} \, =\, \int d\Omega(\bz,\bzbar)\;
\left(Y_{N,k}(\bz)\cdot Y_{N,k}({\bx})\right)
\left(Y_{N,l}(\bz)\cdot Y_{N,l}({\bx})\right)
\left(Y_{N,m}(\bz)\cdot Y_{N,m}({\bx})\right)
\ee
and by using \reff{YYGeg} we end up with
\begin{eqnarray}
&& \hskip -2truecm
   {\cal C}^2_{N;\; k,l,m} = (N-1)
   \, {\cal N}_{N,k}\,  {\cal N}_{N,l} \, {\cal N}_{N,m}\times \nonumber \\
&& \hskip -2truecm 
   \quad \times \int^1_{0} dt\; (1-t)^{N-2}\; \,
   {P^{(N-2,0)}_k (2t-1) \over P^{(N-2,0)}_k (1)} \;
   {P^{(N-2,0)}_l (2t-1) \over P^{(N-2,0)}_l (1)} \;
   {P^{(N-2,0)}_m (2t-1) \over P^{(N-2,0)}_m (1)} \;.
\label{C2integral}
\end{eqnarray}
Using the expansion of $P^{(N-2,0)}_k (x)$ in powers of $(1-x)$ 
(cf.\ \cite{GR}, p.\ 1036) we can write for $|l-m|\le k\le l+m$
\begin{eqnarray}
{\cal C}^2_{N;\; k,l,m} & = &
   (N-1)
   \, {\cal N}_{N,k}\,  {\cal N}_{N,l} \, {\cal N}_{N,m}
\nonumber \\
&& \times \sum_{s=0}^k (-1)^s 
  {k\choose s}\, {(N-2)! \, (N-2+k+s)!\over (N-2+k)!\, (N-2+s)!} \;
  I(s;l,m)\; ,
\end{eqnarray}
where
\be
I(s;l,m)\;=\; 2^{1-N-s} \int_{-1}^1\, dx\; 
   (1-x)^{N-2+s}\; 
   {P^{(N-2,0)}_l (x) \over P^{(N-2,0)}_l (1)} \;
   {P^{(N-2,0)}_m (x) \over P^{(N-2,0)}_m (1)} \;.
\ee
This integral can be evaluated explicitly (cf.\ Formula 2.22.18.2 of 
Ref.\  \cite{Prudnikov}, vol.\ 2). For $s < |l-m|$ we have 
$I(s;l,m) = 0$, while  for $s \ge |l-m|$ we obtain
\begin{eqnarray}
&& \hskip -1.5truecm I(s;l,m) = 
   {l!\, m!\, \left[ (N-2)!\right]^2\over 
    (N-2+l)!\, (N-2+m)!} 
\nonumber \\
&& \hskip -1truecm 
  \times \sum_{t=t_{min}}^m (-1)^{l+t}\,
  {(s+t)!\over t!\, (m-t)!\, (s-l+t)!}\,
  {(N-2+t+m)!\, (N-2+t+s)! \over (N-2+t)! \, (N-1+t+s+l)!}
\end{eqnarray}
where $t_{min} = \max(0,l-s)$.

Finally, we report the computation of 
${\cal R}_N(l+p,p,r;m,l,k)$, defined in Eq.\ \reff{Racahdef},
for $|l+p-r| \le m \le l + p + r$,
$|l-k| \le m \le l + k$ and $|p-k|\le r\le p + k$. 
If one of these inequalities is not satisfied 
${\cal R}_N(l+p,p,r;m,l,k) = 0$, since ${\cal C}_{N;s,t,u} = 0$ if 
$s > t + u$ or if $s < |t-u|$.

Using Eq.\  \reff{Cintegral}, we rewrite $\,{\cal R}_N(l+p,p,r;m,l,k)\,$
as \begin{eqnarray}
&& \hskip -2truecm 
   \int d\Omega(\bz,\bzbar)\, d\Omega(\bw,\bwbar)\, d\Omega(\bx,\bxbar)
\nonumber \\
&& \hyper{l+p}{(\beta)_{l+p}}{(\alpha)_{l+p}} (\bw)\,
   \hyper{l}{(\delta)_l}{(\gamma)_l} (\bz) \,
   \hyper{p}{(\zeta)_p}{(\epsilon)_p} (\bx) \,
   \Clebsch{l+p,l,p}{(\alpha)_{l+p};(\gamma)_l;(\epsilon)_p}{
                     (\beta)_{l+p};(\delta)_l;(\zeta)_p}
\nonumber \\ [3mm]
&& \times \left( Y_{N,m}(\bw) \cdot Y_{N,m}(\bz)\right)
 \left( Y_{N,k}(\bx) \cdot Y_{N,k}(\bz)\right)
 \left( Y_{N,r}(\bx) \cdot Y_{N,r}(\bw)\right)\; .
\end{eqnarray}
We can now use Eq.\ \reff{eq:CG} to rewrite the above equation as 
\begin{eqnarray}
&& \hskip -2truecm
   {\mu_{N,p}\, \mu_{N,l} \over \mu_{N,p+l} }                          
    \int d\Omega(\bz,\bzbar)\, d\Omega(\bw,\bwbar)\, d\Omega(\bx,\bxbar)
\; \left[ Y_{N,l+p} (\bw) \cdot Y_{N,l} (\bz) \cdot Y_{N,p} (\bx)\right]_l
\nonumber \\ [3mm]
&& \times \left( Y_{N,m}(\bw) \cdot Y_{N,m}(\bz)\right)
 \left( Y_{N,k}(\bx) \cdot Y_{N,k}(\bz)\right)
 \left( Y_{N,r}(\bx) \cdot Y_{N,r}(\bw)\right)\; .
\end{eqnarray}
Now we integrate over $\bx$ using Eqs.\  \reff{Cintegral} and 
\reff{eq:CG}. We obtain
\begin{eqnarray}
&& \hskip -2truecm
  {(N-1)!\, \mu_{N,p}^2\, \mu_{N,l}\, \mu_{N,k}\, \mu_{N,r}
           \over (N-1+p+k+r)!\, \mu_{N,p+l} }                          
    \int d\Omega(\bz,\bzbar)\, d\Omega(\bw,\bwbar)
\nonumber \\
&& \sum_{i=i_{min}}^{i_{max}} {k\choose h} {r\choose i}
          {p\choose j}\, k!\, r!\, p!\, 
   \hyper{l}{(\alpha)_l}{(\beta)_l} (\bz) 
\nonumber \\
&& \times \left[ \hyper{l+p}{(\alpha)_l}{(\beta)_l} (\bw) 
         \cdot Y_{N,k} (\bz) \cdot Y_{N,r} (\bw)\right]_{p-r+i}\,
 \left( Y_{N,m}(\bw) \cdot Y_{N,m}(\bz)\right)
\end{eqnarray}
where $h=p-r+i$, $j=p-k+i$, $i_{max} = \min(k,r,k+r-p)$ and 
      $i_{min} = \max(0,r-p,k-p)$.

At this point we notice that the remaining term corresponds to $B_5$,
cf.\ Eq.\  \reff{B5def}. We have therefore
\begin{eqnarray}
{\cal R}_N(l+p,p,r;m,l,k) 
&=& {(N-1)!\, \mu_{N,p}^2\, \mu_{N,l}\, \mu_{N,k}\, \mu_{N,r}
           \over (N-1+p+k+r)!\, \mu_{N,p+l} }                          
\nonumber \\
&& \hskip -2.5truecm \times \,
      \sum_{i=i_{min}}^{i_{max}} {k\choose h} {r\choose i}
          {p\choose j}\, k!\, r!\, p!\, 
  B_5(p-r+i;l;p,k,r,m)\; .
\label{Rrisultato}
\end{eqnarray}

\section*{Acknowledgments}

We thank Alan Sokal for useful discussions.

%
%

\end{document}